\documentclass[aps,pre]{revtex4}
\usepackage[T1]{fontenc}
\usepackage{graphicx}
\usepackage{amssymb,amsfonts,amsmath}
\usepackage{dsfont}
\DeclareMathAlphabet{\varmathbb}{U}{bbold}{m}{n}
\newcommand{\id}{\mathds{1}}

\newcommand{\abs}[1]{\left| #1 \right| }

\DeclareMathOperator*{\tr}{\mathrm{tr}}

\newcommand{\cin}{c_{\mathrm{in}}}
\newcommand{\cout}{c_{\mathrm{out}}}
\newcommand{\tc}{\tilde{c}}
\newcommand{\tcin}{\tilde{c}_{\mathrm{in}}}
\newcommand{\tcout}{\tilde{c}_{\mathrm{out}}}
\newcommand{\cavg}{c}
\newcommand{\inner}[2]{\langle #1, #2 \rangle}
\newcommand{\Exp}{\mathbb{E}}
\newcommand{\R}{\mathbb{R}}
\newcommand{\e}{\mathrm{e}}
\newcommand{\deltain}{\delta^\mathrm{in}}
\newcommand{\deltaout}{\delta^\mathrm{out}}

\begin{document}
\title{Spectral redemption: clustering sparse networks}

\author{
Florent Krzakala$^{1}$, Cristopher Moore$^2$, Elchanan Mossel$^3$\\
Joe Neeman$^3$, Allan Sly$^3$, Lenka Zdeborov\'a$^4$ and Pan Zhang$^{1,2}$\\}

\vspace{0.5cm}

\affiliation{
$^1$ ESPCI and CNRS UMR 7083, 10 rue Vauquelin,Paris  75005\\
$^2$ Santa Fe Institute, 1399 Hyde Park Road, Santa Fe NM 87501, USA\\
$^3$ University of California, Berkeley\\
$^4$ Institut de Physique Th\'eorique, CEA Saclay and URA 2306, CNRS, 91191 Gif-sur-Yvette, France}

\begin{abstract}
Spectral algorithms are classic approaches to clustering and community detection in networks.  However, for sparse networks the standard versions of these algorithms are suboptimal, in some cases completely failing to detect communities even when other algorithms such as belief propagation can do so.  Here we introduce a new class of spectral algorithms based on a non-backtracking walk on the directed edges of the graph.  
The spectrum of this operator is much better-behaved than that of the
adjacency matrix or other commonly used matrices, maintaining a strong separation between the bulk eigenvalues and the eigenvalues relevant to community structure even in the sparse case.  We show that our algorithm is optimal for graphs generated by the stochastic block model, 
detecting communities all the way down to the theoretical limit.  We
also show the spectrum of the non-backtracking operator for some
real-world networks, illustrating its advantages over traditional spectral clustering. 
\end{abstract}

\date{\today}
\maketitle

Detecting communities or modules is a central task in the study of
social, biological, and technological networks.  Two of the most
popular approaches are statistical inference, where we fix a
generative model such as the stochastic block model to the
network~\cite{blockmodel1,blockmodel2}; and spectral methods, where we
classify vertices according to the eigenvectors of a matrix associated
with the network such as its adjacency matrix or
Laplacian~\cite{clustering-intro}.

Both statistical inference and spectral methods have been shown to work well in networks that
are sufficiently dense, or when the graph is regular~\cite{mcsherry,bickel-chen,CoMoVi:09,CO:10,nadakuditi-newman1}.  
However, for sparse networks 
with widely varying degrees, 
the community detection problem is harder.  Indeed, it was recently
shown~\cite{decelle-etal1,decelle-etal2,mossel-neeman-sly} that there
is a phase transition below which communities present in the
underlying block model are impossible for any algorithm to detect.
While standard spectral algorithms succeed down to this transition
when the network is sufficiently dense, with an average degree growing
as a function of network size~\cite{nadakuditi-newman1}, in the case
where the average degree is constant these methods fail
significantly above the transition~\cite{zhang-etal}.  Thus there is a
large regime in which statistical inference succeeds in detecting communities, but where current spectral algorithms fail.

It was conjectured in~\cite{mossel-neeman-sly}  that this gap is artificial and that there exists a spectral algorithm that succeeds all the way to the detectability transition even in the sparse case. Here, we 
propose 
an algorithm based on a linear operator considerably different from the adjacency matrix or its variants: namely, a matrix that represents a walk on the directed edges of the network, with backtracking prohibited. We give strong evidence that this algorithm indeed closes the gap. 

The fact that this operator has better
spectral properties than, for instance, the standard random walk
operator has been used in the past in the context of random matrices and random graphs~\cite{McKay:81,Sodin:07,Friedman:08}. 
In the theory of zeta functions of graphs, it is known as 
the edge adjacency operator, or the Hashimoto matrix~\cite{Hashimoto:89}.  It has been used to
show fast mixing for the non-backtracking random
walk~\cite{AlBeLuSo:07}, and arises in connection to belief propagation~\cite{Watanabe10,Vontobel10}, 
in particular to rigorously analyze the behavior of belief propagation for
clustering problems on regular graphs~\cite{CoMoVi:09}. 
It has also been used as a feature vector to classify graphs~\cite{RenWilson}.  
However, using this operator as a foundation for spectral clustering and community detection appears to be novel.  

We show that the resulting spectral algorithms are optimal for networks generated by the stochastic block model, 
finding communities all the way down to the detectability transition.
That is, at any point above this transition, there is a gap between
the eigenvalues related to the community structure and the bulk
distribution of eigenvalues coming from the random graph structure,
allowing us to find a labeling correlated with the true communities.
In addition to our analytic results on stochastic block models, we
also illustrate the advantages of the non-backtracking operator over existing approaches for some real networks.

\section{Spectral Clustering and Sparse Networks}

\begin{figure}[!t]
\begin{center}
\includegraphics[width=0.5\columnwidth]{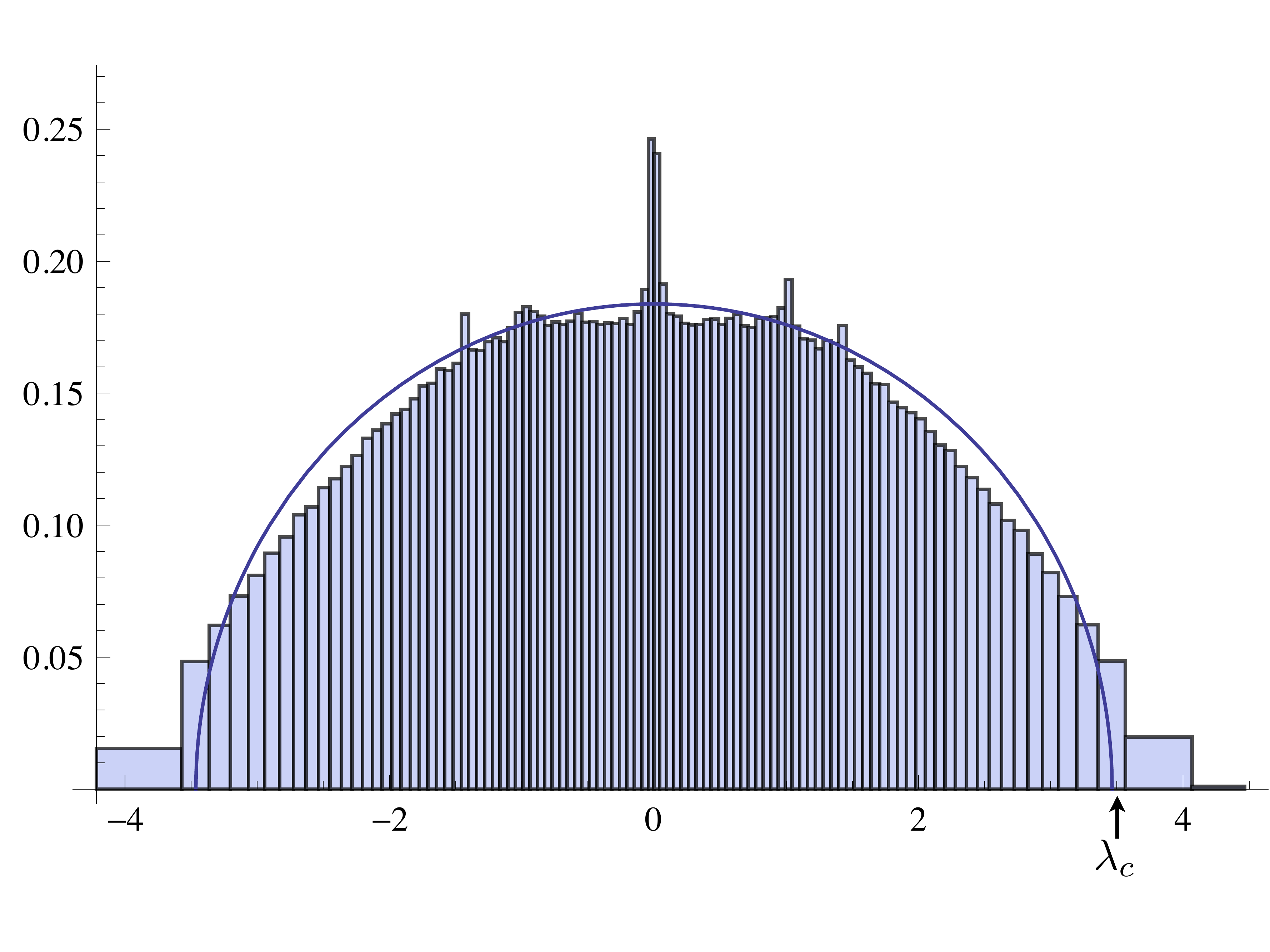}
\end{center}
\caption{The spectrum of the adjacency matrix of a sparse network generated by the block model (excluding the zero eigenvalues).  Here $n=4000$,
 $\cin = 5$, and $\cout = 1$, and we average over $20$ realizations.
 Even though the eigenvalue $\lambda_c = 3.5$ given
 by~\eqref{eq:lambda2} satisfies the threshold condition~\eqref{eq:threshold} and lies outside
 the semicircle of radius $2\sqrt{c}=3.46$, deviations from the semicircle law cause it to get
 lost in the bulk, and the eigenvector of the second largest
 eigenvalue is uncorrelated with the community structure.  As a
 result, spectral algorithms based on $A$ are unable to identify the
 communities in this case. \label{fig:confused}} 
\end{figure}
In order to study the effectiveness of spectral algorithms in a
specific ensemble of graphs, suppose that a graph $G$ is generated by the stochastic block model~\cite{blockmodel1}.  There are $q$ groups of vertices, and each vertex $v$ has a group label $g_v \in \{1,\ldots,q\}$.  Edges are generated independently according to a $q \times q$ matrix $p$ of probabilities, with $\Pr[A_{u,v} = 1] = p_{g_u,g_v}$.  In the sparse case, we have $p_{ab} = c_{ab}/n$, where the affinity matrix $c_{ab}$ stays constant in the limit $n \to \infty$.

For simplicity we first discuss the commonly-studied case where $c$
has two distinct entries, $c_{ab} = \cin$ if $a = b$ and $\cout$ if $a
\ne b$.  We take $q=2$ with two groups of equal size, and assume
that the network is assortative, i.e., $\cin > \cout$.  We
summarize the general case of more groups, arbitrary degree
distributions, and so on in subsequent sections below.

The group labels are hidden from us, and our goal is to infer them from the graph.  Let $\cavg = (\cin+\cout)/2$ denote the average degree.  The detectability threshold~\cite{decelle-etal1,decelle-etal2,mossel-neeman-sly} states that in the limit $n \to \infty$, unless
\begin{equation}
\label{eq:threshold}
\cin - \cout > 2 \sqrt{c} \, ,
\end{equation}
the randomness in the graph washes out the block structure to the extent that 
no algorithm can label the vertices better than chance.
Moreover, \cite{mossel-neeman-sly} proved that below this threshold, it is impossible to identify the parameters 
$\cin$ and $\cout$, while above the threshold the parameters $\cin$ and $\cout$ are easily identifiable.

The adjacency matrix is defined as the $n \times n$ matrix $A_{u,v} = 1$ if $(u,v) \in E$ and $0$ otherwise.  A typical spectral algorithm assigns each vertex a $k$-dimensional vector according to its entries in the first $k$ eigenvectors of $A$ for some $k$, and clusters these vectors according to a heuristic such as the $k$-means algorithm (often after normalizing or weighting them in some way).  In the case $q=2$, we can simply label the vertices according to the sign of the second eigenvector.  

As shown in~\cite{nadakuditi-newman1}, spectral algorithms succeed all
the way down to the threshold~\eqref{eq:threshold} if the graph is
sufficiently dense. In that case $A$'s spectrum has a discrete part and a continuous part in the limit $n \to \infty$.  
Its first eigenvector essentially sorts vertices according to their degree,
while the second eigenvector is correlated with the communities. 
The second eigenvalue is given by 
\begin{equation}
\label{eq:lambda2}
\lambda_c = \frac{\cin-\cout}{2} + \frac{\cin+\cout}{\cin-\cout} \, .
\end{equation}
The question is when this eigenvalue gets lost in the continuous bulk
of eigenvalues coming from the randomness in the graph.  This part of
the spectrum, like that of a sufficiently dense Erd\H{o}s-R\'enyi random graph, is asymptotically distributed according to Wigner's semicircle law~\cite{Wigner}
\[
P(\lambda) = \frac{1}{2 \pi \cavg} \,\sqrt{4 \cavg - \lambda^2} \, . 
\]
Thus the bulk of the spectrum lies in the interval $[-2\sqrt{\cavg},2\sqrt{\cavg}]$.  If $\lambda_c > \cavg$, which is equivalent to~\eqref{eq:threshold}, the spectral algorithm can find the corresponding eigenvector, and it is correlated with the true community structure.

However, in the sparse case where $\cavg$ is constant while $n$ is large, this picture breaks down due to a number of reasons.  Most importantly, the leading eigenvalues of $A$ are dictated by 
the vertices of highest degree, 
and the corresponding eigenvectors are localized around these vertices~\cite{KrivelevichSudakov:03}.  As $n$ grows, these eigenvalues exceed $\lambda_c$, 
swamping the community-correlated eigenvector, if any, with the bulk of uninformative eigenvectors.  
As a result, spectral algorithms based on $A$ fail a significant distance from the threshold given by~\eqref{eq:threshold}.  Moreover, this gap grows as $n$ increases: for instance, the largest eigenvalue grows as the square root of the largest degree, which is roughly proportional to $\log n / \log \log n$ for Erd\H{o}s-R\'enyi graphs.  To illustrate this problem, the spectrum of $A$ for a large graph generated by the block model is depicted in Fig.~\ref{fig:confused}. 

Other popular operators for spectral clustering include the Laplacian $L = D - A$ where $D_{uv} = d_u \delta_{u,v}$ is the diagonal matrix of vertex degrees, the random walk matrix $Q_{uv} = A_{uv} / d_u$, and the modularity matrix $M_{uv} = A_{uv} - d_u d_v / (2m)$.  However, all these experience qualitatively the same difficulties as with $A$ in the sparse case.  
Another simple heuristic is to simply remove the high-degree vertices (e.g.~\cite{CO:10}), but this throws away a significant amount of information; in the sparse case it can even destroy the giant component, causing the graph to fall apart into disconnected pieces~\cite{BoJaRi:07}.

\section{The Non-Backtracking Operator}
\begin{figure}[!t]
\begin{center}
\includegraphics[width=0.5\columnwidth]{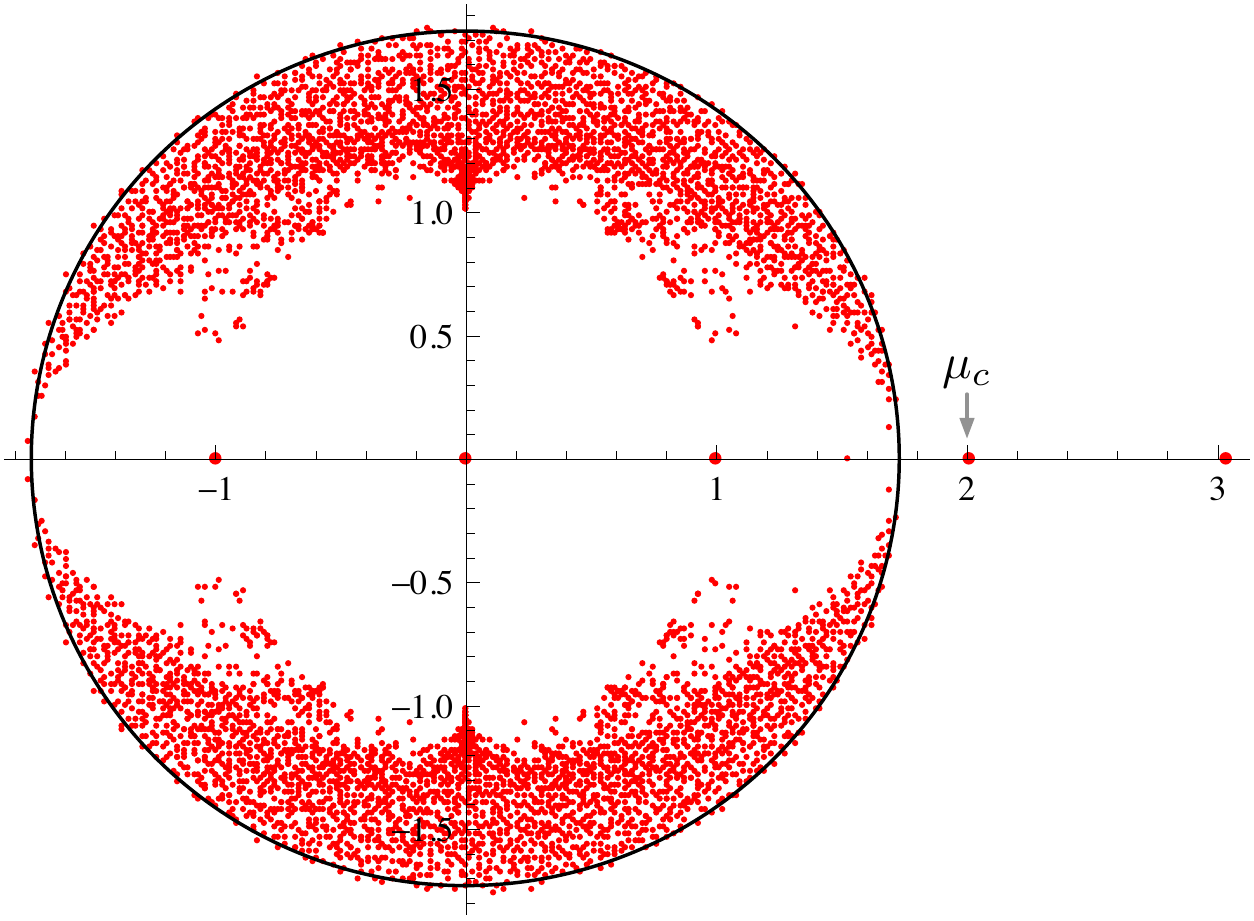} 
\caption{The spectrum of the non-backtracking matrix $B$ for a network generated 
by the block model with same parameters as in Fig.~\ref{fig:confused}.  The leading eigenvalue is at
  $c = 3$, the second eigenvalue is close to $\mu_c = (\cin-\cout)/2 = 2$,
  and the bulk of the spectrum is confined to the disk of radius
  $\sqrt{c} = \sqrt{3}$.  
  Since $\mu_c$ is outside the bulk, a spectral
  algorithm that labels vertices according to 
  the sign of $B$'s second eigenvector (summed over the incoming edges at each vertex) 
  labels the majority of vertices correctly. 
\label{fig:complexplot}}
\end{center}
\end{figure}

The main contribution of this paper is to show how to redeem the performance of spectral algorithms in sparse networks by using a different linear operator.  The \emph{non-backtracking matrix} $B$ is a $2m \times 2m$ matrix, defined on the directed edges of the graph.  Specifically, 
\[
B_{(u \to v),(w \to x)} = \begin{cases} 
1 & \mbox{if $v=w$ and $u \ne x$} \\
0 & \mbox{otherwise} \, . 
\end{cases}
\]
Using $B$ rather than $A$ addresses the problem described above.
The spectrum of $B$ is not sensitive to high-degree vertices, since a walk
starting at $v$ cannot turn around and return to it immediately.
Other convenient properties of $B$ are that any tree dangling off the graph, or disconnected from it, 
simply contributes zero eigenvalues to the spectrum, since a non-backtracking walk is forced to a leaf of the tree where it has nowhere to go.  
Similarly one can show that unicyclic components yield eigenvalues that are either $0$, $1$ or $-1$.

\begin{figure}[t]
\begin{center}
\includegraphics[width=0.4\columnwidth]{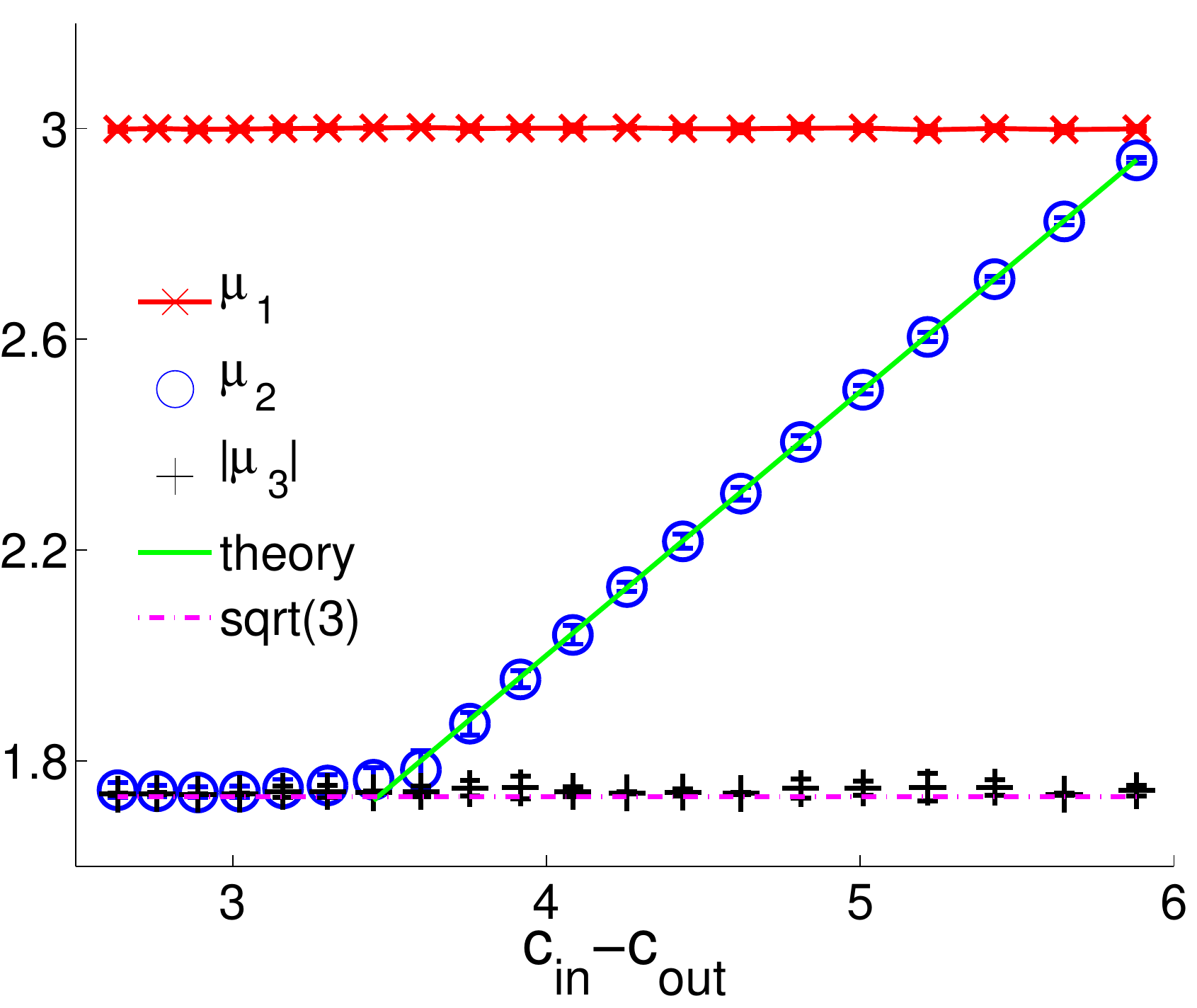}
\end{center}
\caption{The first, second and third largest eigenvalues $\mu_1$, $\mu_2$ and $|\mu_3|$ respectively of $B$ as functions of $\cin-\cout$. The third eigenvalue is complex, so we plot its modulus. Values are averaged over $20$ networks of size $n=10^5$ and average degree $c=3$. The green line in the figure represents $\mu_c=(\cin-\cout)/2$, and the horizontal lines are $c$ and $\sqrt{c}$ respectively.  The second eigenvalue $\mu_2$ is well-separated from the bulk throughout the detectable regime.\label{fig:c3d2}}
\end{figure}

As a result, $B$ has the following spectral properties in the limit
$n\!\to\!\infty$ 
in the ensemble of graphs generated by the block model. 
The leading eigenvalue is the average degree 
$\cavg = (\cin+\cout)/2$.  At any point above the detectability
threshold~\eqref{eq:threshold}, the second eigenvalue 
is associated with the block structure and reads
\begin{equation}
\label{eq:mu_q2}
\mu_c = \frac{\cin-\cout}{2}\, .  
\end{equation}
Moreover, the bulk of $B$'s spectrum is confined to the disk in the
complex plane of radius $\sqrt{c}$, as shown in
Fig.~\ref{fig:complexplot}.  
As a result, the second eigenvalue is well separated from the top of the bulk, i.e., from the third largest eigenvalue in absolute value, as shown in 
Fig.~\ref{fig:c3d2}.

The eigenvector corresponding to $\mu_c$ is strongly correlated with the community structure.  
Since $B$ is defined on directed edges, at each vertex we sum this eigenvector over all its incoming edges.  If we label vertices according to the sign of this sum, then the majority of vertices are labeled correctly (up to a change of sign, which switches the two communities).  Thus a spectral algorithm based on $B$ succeeds when $\mu_c > \sqrt{c}$, 
i.e. when~\eqref{eq:threshold} holds---but unlike standard spectral algorithms, this criterion now holds even in the sparse case.  
We present arguments for these claims in the next section.

\section{Reconstruction and a Community-Correlated Eigenvector}

In this section we sketch justifications of the claims in the previous section regarding $B$'s spectral properties, showing that its second eigenvector is correlated with the communities whenever~\eqref{eq:threshold} holds.  
Let us start by recalling how to generalize
equation~\eqref{eq:lambda2} for the adjacency matrix $A$ of sparse graphs. 
We follow~\cite{mossel-neeman-sly}, who derived a similar result in the case of random regular graphs.

With $\mu=\mu_c$ defined as in~\eqref{eq:mu_q2},   
for a given integer $r$, consider the vector 
\begin{equation}
f^{(r)}_v = \mu^{-r} \sum_{u : d(u,v) = r} \sigma_u \, , 
\end{equation}
where $\sigma_u = \pm 1$ denotes $u$'s community.  By the theory of the reconstruction problem on trees~\cite{KestenStigum:66,MosselPeres:03}, if~\eqref{eq:threshold} holds then the correlation $\inner{f^{(r)}}{\sigma}/n$ is bounded away from zero in the limit $n \to \infty$.

We will show that if $r$ is large but small compared to the diameter of the graph, then $f^{(r)}$ is closely related to the second eigenvector of $B$.  Thus if we label vertices according to the sign of this second eigenvector (summed over all incoming edges at each vertex) we obtain the true communities with significant accuracy.

First we show that $f^{(r)}$ approximately obeys an eigenvalue equation that generalizes~\eqref{eq:lambda2}.  As long as the radius-$r$ neighborhood of $v$ is a tree, we have 
\begin{align*}
(Af^{(r)})_v 
&= \mu^{-r} \left[ \sum_{u: d(u,v) = r+1} \sigma_u + (d_v-1) \sum_{u: d(u,v) = r-1} \sigma_u \right] \, ,
\end{align*}
so 
\begin{equation}
\label{eq:approx-A}
(A f^{(r)})_v = \mu f^{(r+1)}_v + (d_v - 1) \mu^{-1} f^{(r-1)}_v \, . 
\end{equation}
Summing over $v$'s neighborhood gives the expectation 
\[
\Exp \Bigg[ \sum_{u \in N(v)} \sigma_u \Bigg] = \mu \sigma_v \, , 
\]
and summing the fluctuations over the (in expectation) $c^r$ vertices at distance $r$ gives 
\[
\abs{ f^{(r)}_v - f^{(r \pm 1)}_v } = O(c^{r/2} \mu^{-r}) \, .
\]
If $\mu=\mu_c$ and~\eqref{eq:threshold} holds so that $\mu_c > \sqrt{c}$, these fluctuations tend to zero for large $r$.  In that case, we can identify $f^{(r)}$ with $f^{(r \pm 1)}$, and \eqref{eq:approx-A} becomes
\begin{equation}
\label{eq:quad_sym}
A f = \mu f + (D - \id) \mu^{-1} f \, . 
\end{equation}
In particular, in the dense case we can recover~\eqref{eq:lambda2} by approximating $D$ with $c \id$, 
or equivalently pretending that the graph is $c$-regular.  Then $f$ is an eigenvector of $A$ with eigenvalue $\lambda_c = \mu + (c-1) \mu^{-1}$.

We define an analogous approximate eigenvector of $B$, 
\[
g^{(r)}_{u \to v} = \mu^{-r} \sum_{(w,x) : d(u \to v, w \to x) = r} \sigma_x \, ,
\]
where now $d$ refers to the number of steps in the graph of directed edges.  We have in expectation 
\[
B g^{(r)} = \mu g^{(r+1)} \, , 
\]
and as before $|g^{(r)}-g^{(r+1)}|$ tends to zero as $r$ increases.  Identifying them gives an approximate eigenvector $g$ with eigenvalue $\mu$, 
\begin{equation}
\label{eq:simple_eigen}
B g = \mu g \, . 
\end{equation}
Furthermore, summing over all incoming edges gives 
\[
\sum_{u \in N(v)} g_{u \to v} = f_v \, ,
\]
giving signs correlated with the true community memberships $\sigma_v$.

We note that the relation between the eigenvalue
equation~\eqref{eq:simple_eigen} for $B$ and the quadratic eigenvalue
equation~\eqref{eq:quad_sym} is exact and well known in the theory of
zeta functions of graphs \cite{Hashimoto:89,Bass92,AnFrHo:07}.  
More generally, all eigenvalues $\mu$ of $B$ that are not $\pm 1$ are the roots of the equation
\begin{equation} 
\label{eq:ihara}
\det \left[ \mu^2 \id - \mu A + (D-\id) \right] = 0 \, . 
\end{equation} 
This equation hence describes $2n$ of $B$'s eigenvalues.  
These are  the eigenvalues of a $2n \times 2n$ matrix, 
\begin{equation}
\label{eq:2nby2n}
B' = \begin{pmatrix}
0 & D-\id \\
-\id & A
\end{pmatrix} \, .
\end{equation}
The left eigenvectors of $B'$ are of the form $(f, -\mu f)$ where $f$ obeys~\eqref{eq:quad_sym}.  Thus we can find $f$ by dealing with a $2n \times 2n$ matrix rather than a $2m \times 2m$ one, which considerably reduces the computational complexity of our algorithm.

Next, we argue that the bulk of $B$'s spectrum is confined to the disk of radius $\sqrt{c}$.  First note that for any matrix $B$, 
\[
\sum_{i=1}^{2m} \abs{\mu_i}^{2r} \le \tr B^r (B^r)^T \, .
\]
On the other hand, for any fixed $r$, since $G$ is locally treelike in the limit $n \to \infty$, each diagonal entry $(u \to v, u \to v)$ of $B^r (B^r)^T$ is equal to the number of vertices exactly $r$ steps from $v$, other than those connected via $u$.  In expectation this is $c^r$, so by linearity of expectation $\Exp \tr B^r (B^r)^T = 2m c^r$.   In that case, the spectral measure has the property that 
 \[
\Exp (|\mu|^{2r}) \le c^r \, . 
\]
Since this holds for any fixed $r$, we conclude that almost all of $B$'s eigenvalues obey $\abs{\mu} \le \sqrt{c}$.  
Proving rigorously that \emph{all} the eigenvalues in the bulk are asymptotically
confined to this disk requires a more precise argument and is left for future work.

As a side remark we note that~\eqref{eq:ihara} yields 
$B$'s spectrum for $d$-regular graphs~\cite{AnFrHo:07}.  There are $n$ pairs of eigenvalues $\mu_\pm$
such that
\begin{equation} 
\label{eq:ihara_reg}
    \mu_\pm = \frac{\lambda \pm \sqrt{\lambda^2 -4(d-1)}}{2} \, , 
\end{equation} 
where $\lambda$ are the (real) eigenvalues of $A$.  These are related by $\mu_+ \mu_- = d-1$, so all the non-real eigenvalues of $B$ are conjugate pairs on the circle of radius $\sqrt{d-1}$.  The other eigenvalues are $\pm 1$.  
For random regular graphs, the asymptotic spectral density of $B$ follows straightforwardly from the well known result of~\cite{McKay:81}   
for the spectral density of the adjacency matrix.

Finally, the \emph{singular} values of $B$ are easy to derive for any simple graph, i.e., one without self-loops or multiple edges.  Namely, $B B^T$ is block-diagonal: for each vertex $v$, it has a rank-one block of size $d_v$ that connects $v$'s outgoing edges to each other.  As a consequence, $B$ has $n$ singular values $d_v-1$, and its other $2m-n$ singular values are $1$.  However, since $B$ is not symmetric, its eigenvalues and its singular values are different---while its singular values are controlled by the vertex degrees, its eigenvalues are not.  
This is precisely why its spectral properties are better than those of $A$ and related operators.

\section{More Than Two Groups and General Degree Distributions}

The arguments given above regarding $B$'s spectral properties generalize straightforwardly to other graph ensembles. 
First, consider block models with $q$ groups, where for $1 \le a \le q$ group $a$ has fractional size $n_a$.  The average degree of group $a$ is $c_a = \sum_b c_{ab} n_b$.  The hardest case is where $c_a=c$ is the same for all $a$, so that we cannot simply label vertices according to their degree.  

The leading eigenvector again has eigenvalue $c$, and the bulk of $B$'s spectrum is again confined to the disk of radius $\sqrt{c}$.  Now $B$ has $q-1$ linearly independent eigenvectors with real eigenvalues, and the corresponding eigenvectors are correlated with the true group assignment.  If these real eigenvalues lie outside the bulk, we can identify the groups by assigning a vector in $\R^{q-1}$ to each vertex, and applying a clustering technique such as $k$-means.  These eigenvalues are of the form $\mu = c \nu$ where $\nu$ is a nonzero eigenvalue of the $q \times q$ matrix
\begin{equation}
\label{eq:T}
     T_{ab} = n_a \left( \frac{c_{ab}}{c}-1 \right) \, . 
\end{equation}
In particular, if $n_a = 1/q$ for all $a$, and $c_{ab} = \cin$ for $a=b$ and $\cout$ for $a \ne b$, we have $\mu_c = (\cin-\cout)/q$.  The detectability threshold is again $\mu_c > \sqrt{c}$, or 
\begin{equation}
\label{eq:threshold-gen}
\abs{\cin - \cout} > q \sqrt{c} \, .
\end{equation}
More generally, if the community-correlated eigenvectors have distinct
eigenvalues, we can have multiple transitions where some of them can
be detected by a spectral algorithm while others cannot.

There is an important difference between the general case and $q=2$.
While for $q=2$ it is literally impossible for any algorithm
to distinguish the communities below this transition, for larger $q$
the situation is more complicated.  In general (for $q \ge 5$ in the
assortative case, and $q \ge 3$ in the disassortative one) the
threshold~\eqref{eq:threshold-gen} marks a transition from an ``easily
detectable'' regime to a ``hard detectable'' one.  In the hard
detectable regime, it is theoretically possible to find the
communities, but it is conjectured that any algorithm that does so
takes exponential time~\cite{decelle-etal1,decelle-etal2}.  In
particular, we have found experimentally that none of $B$'s
eigenvectors are correlated with the groups in the hard regime.
Nonetheless, our arguments suggest that spectral algorithms based on
$B$ are optimal in the sense that they succeed all the way down to
this easy/hard transition.  

Since a major drawback of the stochastic block model is that its
degree distribution is Poisson, we can also consider random graphs
with specified degree distributions.  Again, the hardest case is where
the groups have the same degree distribution.  
Let $a_k$ denote the fraction of vertices of degree $k$.  The 
average branching ratio of a branching process that explores the neighborhood of a vertex, i.e., the average number of new edges leaving a vertex $v$ that we arrive at when following a random edge, is
\[
\tc 
= \frac{\sum_k k(k-1) a_k}{\sum_k k a_k} 
= \langle k^2 \rangle / \langle k \rangle - 1 \, . 
\]
We assume here that the degree distribution has bounded second moment so that this process is not dominated by a few high-degree vertices.  The leading eigenvalue of $B$ is $\tc$, and the bulk of its spectrum is confined to the disk of radius $\sqrt{\tc}$, even in the sparse case where $\tc$ does not grow with the size of the graph.  If $q=2$ and the average numbers of new edges linking $v$ to its own group and the other group are $\tcin/2$ and $\tcout/2$ respectively, then the approximate eigenvector described in the previous section has eigenvalue $\mu = (\tcin-\tcout)/2$.  The detectability threshold~\eqref{eq:threshold} then becomes $\mu > \sqrt{\tc}$, or $\tcin - \tcout > 2 \sqrt{\tc}$.  The threshold~\eqref{eq:threshold-gen} for $q$ groups generalizes similarly.

\section{Deriving B by Linearizing Belief Propagation}

The matrix $B$ also appears naturally as a linearization of the update equations for belief propagation (BP).  This linearization was used previously to investigate phase transitions in the performance of the BP algorithm~\cite{CoMoVi:09,Urbanke,decelle-etal1,decelle-etal2}. 
 
We recall that BP is an algorithm that iteratively updates messages
$\eta_{v \to w}$ where $(v,w)$ are directed edges.  
These messages represent the marginal probability that a vertex $v$ belongs to a given community, 
assuming that the vertex $w$ is absent from the network.  
Each such message is updated according to the messages $\eta_{u \to v}$ that
$v$ receives from its other neighbors $u \ne w$.  The update rule
depends on the parameters $\cin$ and $\cout$ of the block model, as
well as the expected size of each community.  For the simplest case of
two equally sized groups, the BP update~\cite{decelle-etal1,decelle-etal2} can be written as
\begin{equation}
\label{eq:bp-update}
 \frac{\eta^+_{v \to w}}{\eta^-_{v \to w}} := 
\e^{-h} \frac{\prod_{u \in N(v) - w} \big( \eta^+_{u \to w} \cin + \eta^-_{u \to w} \cout \big)}
 {\prod_{u \in N(v) - w} \big( \eta^+_{u \to w} \cout + \eta^-_{u \to w} \cin \big)} \, . 
\end{equation}
Here $+$ and $-$ denote the two communities.  
The term $\e^h$, 
where $h = (\cin - \cout)(n^\mathrm{BP}_+ - n^\mathrm{BP}_-)$ and $n^{\mathrm BP}_\pm$ is the current estimate of the fraction of vertices in the two groups, represents messages from the non-neighbors of $v$.  In the assortative case, it prevents BP from converging to a fixed point where every vertex is in the same community.

The update~\eqref{eq:bp-update} has a trivial fixed point $\eta_{v \to w} = 1/2$, where every vertex is equally likely to be in either community.  
Writing $\eta_{u \to v}^\pm = 1/2 \pm \delta_{u \to v}$ and linearizing around this fixed point gives the following update rule for $\delta$, 
\[
\delta_{v \to w} := \frac{\cin - \cout}{ \cin + \cout } \sum_{u \in N(v) - w} \delta_{u \to v} \, , 
\]
or equivalently
\begin{equation}
\label{eq:delta}
\delta := \frac{\cin - \cout}{ \cin + \cout } B \delta \, . 
\end{equation}

More generally, in a block model with $q$ communities, an affinity matrix $c_{ab}$, and an expected fraction $n_a$ of vertices in
each community $a$, 
linearizing around the trivial point and defining $\eta_{u \to v}^a = n_a + \delta_{u \to v}^a$ gives a tensor product operator
\begin{equation}
\label{eq:tensor}
 \delta := (T \otimes B) \delta \, ,
\end{equation}
where $T$ is the $q \times q$ matrix defined in~\eqref{eq:T}.  

We can also describe the linearization of BP in terms of the $2n \times 2n$ matrix $B'$ defined in~\eqref{eq:2nby2n}.  Specifically, if we define $\deltain$ and $\deltaout$ as the $qn$-dimensional vectors where $\deltain_v = \sum_{u \in N(v)} \delta_{u \to v}^a$ and $\deltaout_v = \sum_{u \in N(v)} \delta_{v \to u}^a$ are the sum of $\delta$ over $v$'s incoming and outgoing edges respectively, then 
\begin{equation}
\label{eq:tensor-2nby2n}
\begin{pmatrix} \deltaout \\ \deltain \end{pmatrix}
= (T \otimes B') 
\begin{pmatrix} \deltaout \\ \deltain \end{pmatrix} \, . 
\end{equation}
Thus we can analyze BP to first order around the trivial fixed point by keeping track of just $2qn$ variables rather than $2qm$ of them.

This shows that the spectral properties of the non-backtracking matrix are closely related to belief propagation. 
Specifically, the trivial fixed point is unstable, leading to a fixed point that is correlated with the community structure, exactly when $T \otimes B$ has an eigenvalue greater than $1$.  However, by avoiding the fixed point where all the vertices belong to the same group, we suppress $B$'s leading eigenvalue; thus the criterion for instability is $\nu \mu_2 > 1$ where $\nu$ is $T$'s leading eigenvalue and $\mu_2$ is $B$'s second eigenvalue.  This is equivalent to~\eqref{eq:threshold-gen} in the case where the groups are of equal size. 

In general, the BP algorithm provides a slightly better agreement with the actual group assignment, since it approximates the Bayes-optimal inference of the block model. On the other hand, the BP update rule depends on the parameters of the block model, and if these parameters are unknown they need to be learned, which presents additional difficulties~\cite{zhang-etal}.  In contrast, our spectral algorithm does not depend on the parameters of the block model, giving an advantage over BP in addition to its computational efficiency.

\section{Experimental Results and Discussion}

\begin{figure}[!ht]
\begin{center}
\includegraphics[width=0.45\columnwidth]{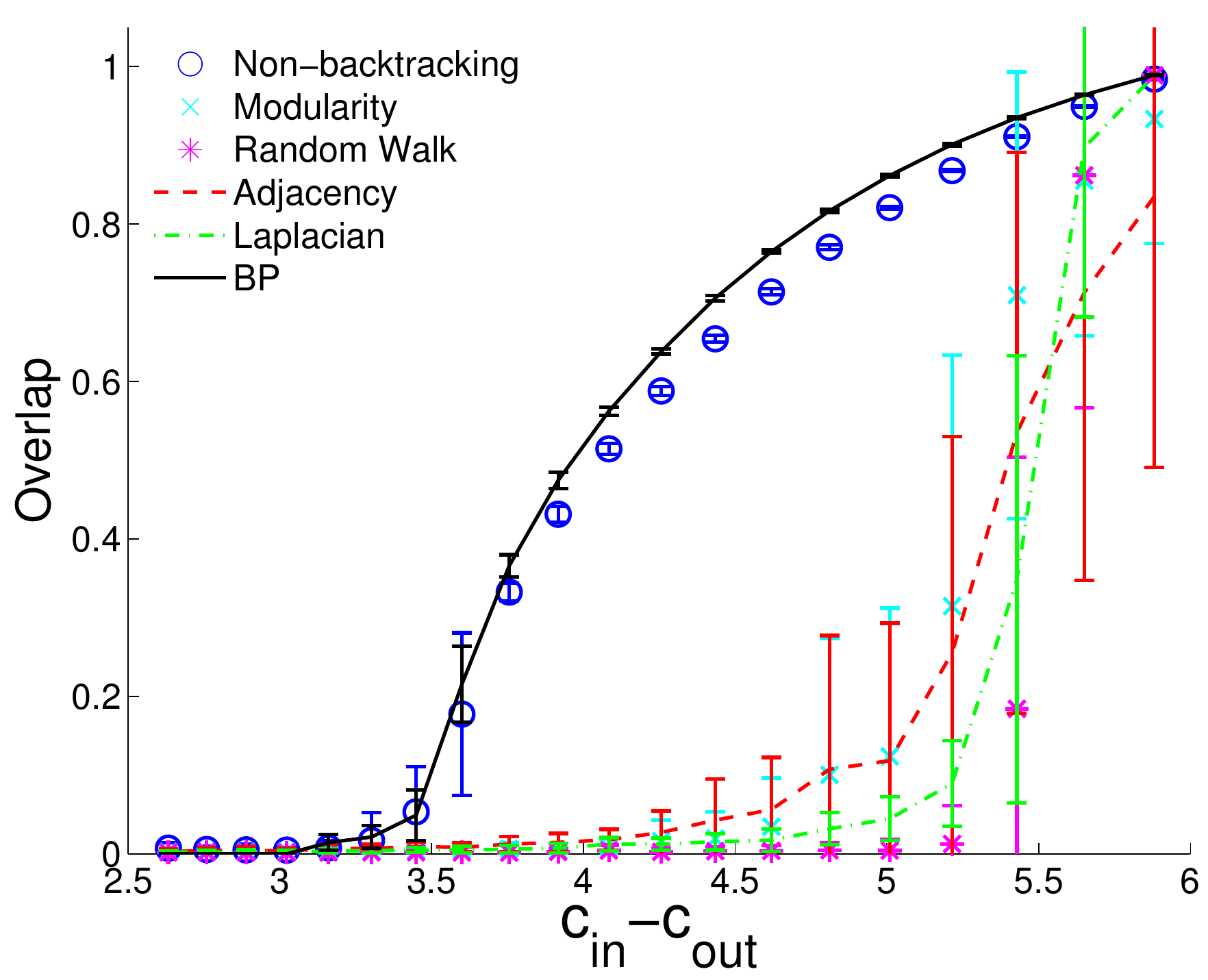}
\includegraphics[width=0.45\columnwidth]{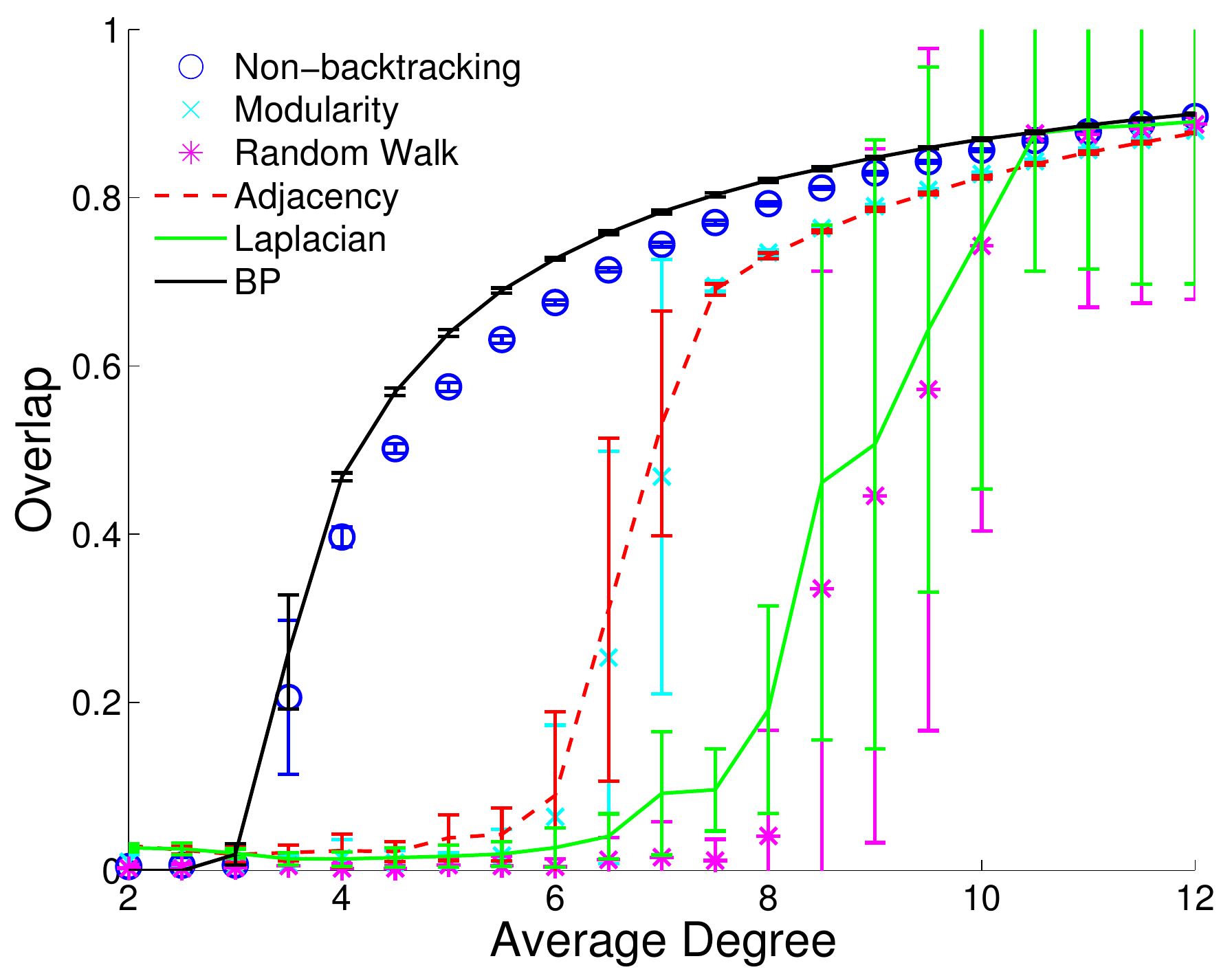}
\end{center}
\caption{The accuracy of spectral algorithms based on different linear operators, and of belief propagation, for two groups of equal size.  On the left, we vary $\cin-\cout$ while fixing the average degree $\cavg=3$; the detectability transition given by~\eqref{eq:threshold} occurs at $\cin-\cout = 2 \sqrt{3} \approx 3.46$.  On the right, we set $\cout/\cin=0.3$ and vary $\cavg$; the detectability transition is at $\cavg \approx 3.45$.  Each point is averaged over $20$ instances with $n=10^5$.  Our spectral algorithm based on the non-backtracking matrix $B$ achieves an accuracy close to that of BP, and both remain large all the way down to the transition.  Standard spectral algorithms based on the adjacency matrix, modularity matrix, the Laplacian, and the random walk matrix fail well above the transition, doing no better than chance.\label{fig:ovl:c3}}
\end{figure}

\begin{figure}[!ht]
\begin{center}
\includegraphics[width=0.3\columnwidth]{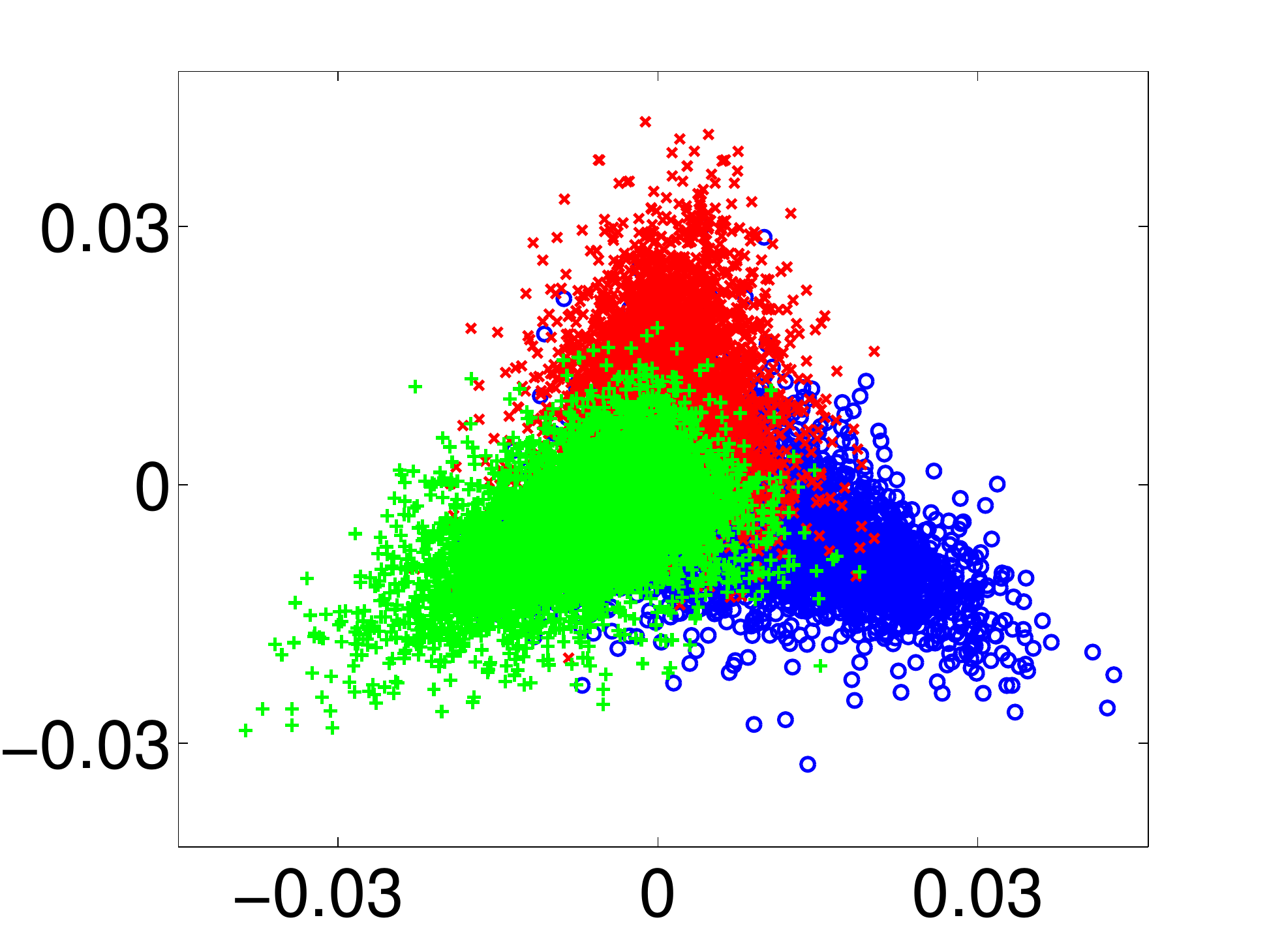}
\includegraphics[width=0.3\columnwidth]{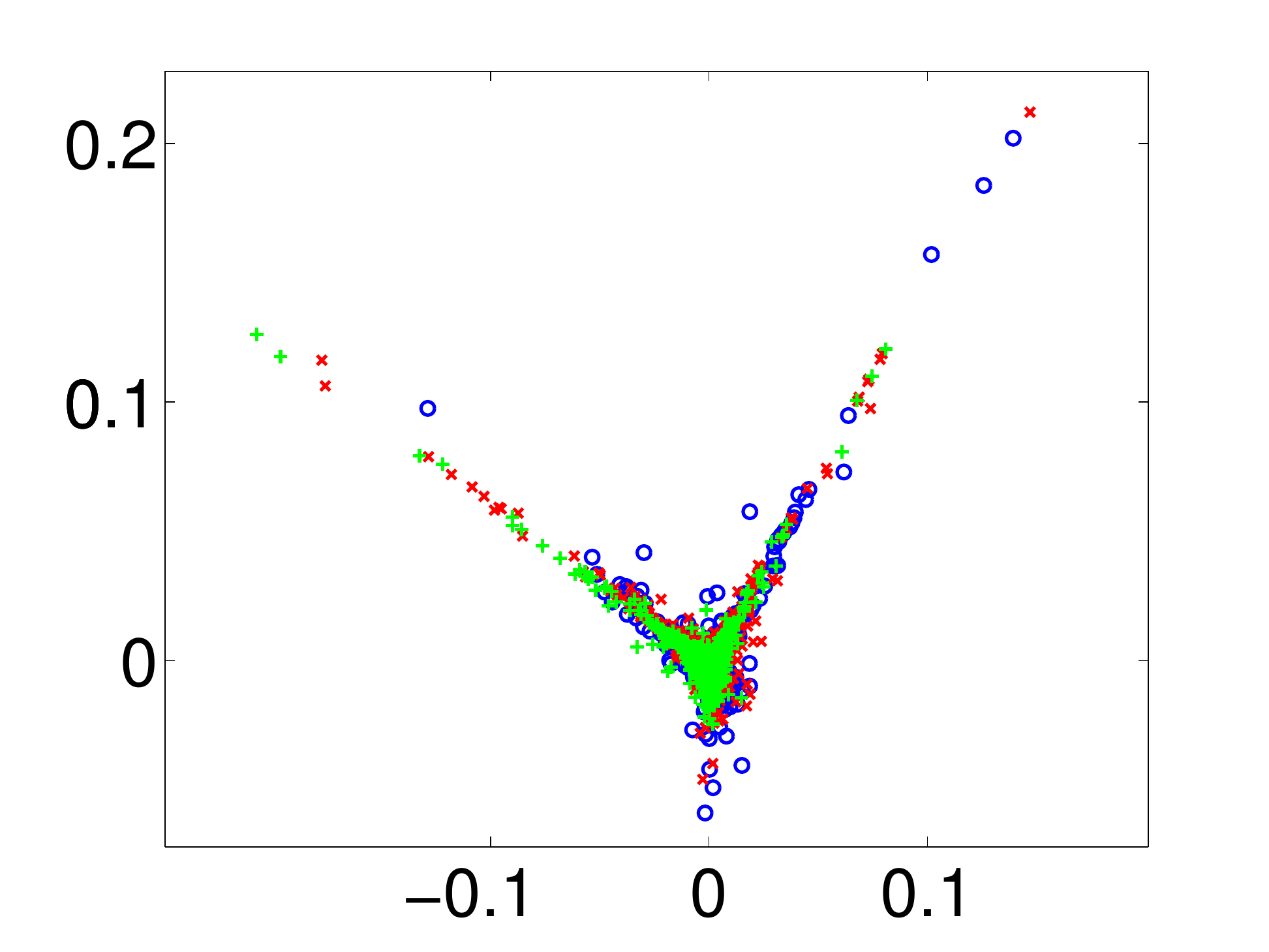}
\end{center}
\caption{Clustering in the case of three groups of equal size.  On the
  left, a scatter plot of the second and third eigenvectors (X and Y
  axis respectively) of the non-backtracking matrix $B$, with colors
  indicating the true group assignment.  On the right, the analogous
  plot for the adjacency matrix $A$.  Here $n=3 \times 10^4$, $c=3$,
  and $\cout / \cin=0.1$.  Applying $k$-means gives an overlap $0.712$
  using $B$, but $0.0063$ using $A$. \label{fig:q3}}
\end{figure}

In Fig.~\ref{fig:ovl:c3}, we compare the spectral algorithm based on the non-backtracking matrix $B$ with those based on various classical operators: the adjacency matrix $A$, the modularity matrix $M$, the Laplacian $L$, and the random walk matrix $Q$.  We see that there is a regime where standard spectral algorithms do no better than chance, while the one based on $B$ achieves a strong correlation with the true group assignment all the way down to the detectability threshold.  We also show the performance of belief propagation, which is believed to be asymptotically optimal~\cite{decelle-etal1,decelle-etal2}.

We measure the performance as the \emph{overlap}, defined as 
\begin{equation}
\label{eq:overlap}
        \left( \frac{1}{n} \sum_u \delta_{g_u , \tilde{g}_u} -
          \frac{1}{q}\right) \left/ \left( 1 -  \frac{1}{q} \right) \right. \, .
\end{equation}
Here $g_u$ is the true group label of vertex $u$, and $\tilde{g}_u$ is the label found by the algorithm.  
We break symmetry by maximizing over all $q!$ permutations of the groups.  The overlap is normalized so that it is $1$ for the true labeling, and $0$ for a uniformly random labeling.

In Fig.~\ref{fig:q3} we illustrate clustering in the case $q=3$.  As described above, in the detectable regime we expect to see $q-1$ eigenvectors with real eigenvalues that are correlated with the true group assignment.  Indeed $B$'s second and third eigenvector are strongly correlated with the true clustering, and applying $k$-means in $\R^2$ gives a large overlap.  In contrast, the second and third eigenvectors of the adjacency matrix are essentially uncorrelated with the true clustering, and similarly for the other traditional operators.

\begin{figure}[!t]
\begin{center}
\includegraphics[width=\columnwidth]{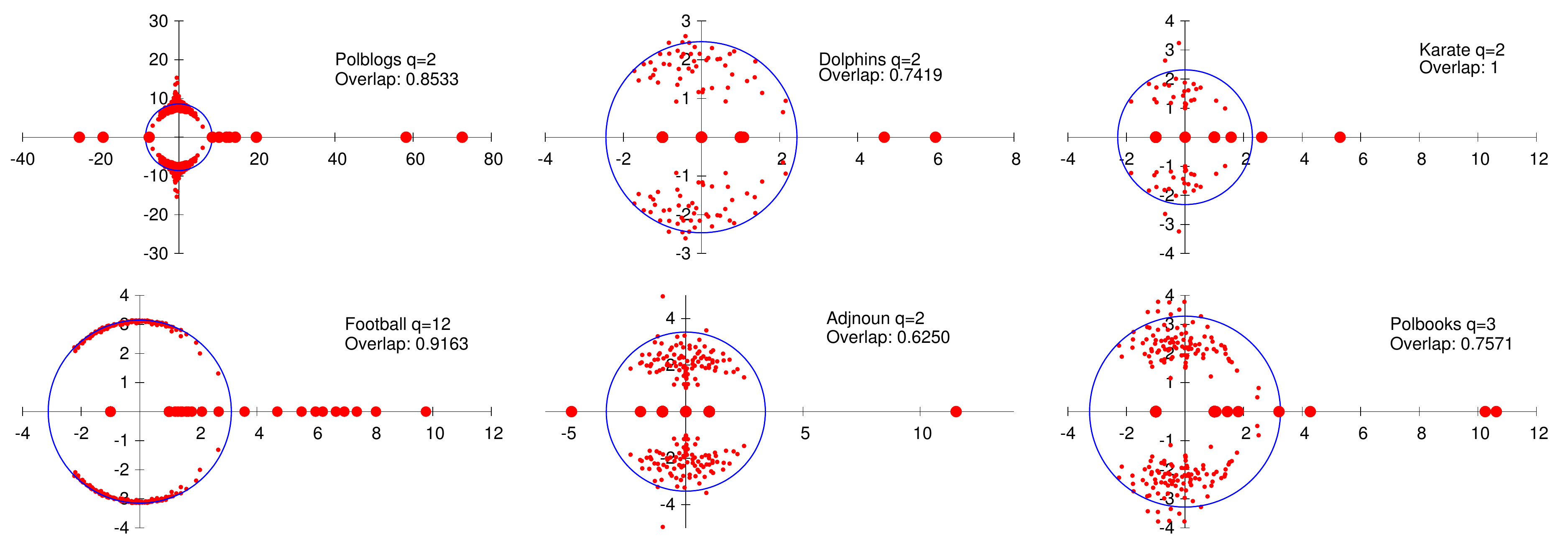}
\end{center}
\caption{Spectrum of the non-backtracking matrix in the complex plane for some commonly used benchmarks for community detection in real networks taken from~\cite{lada,zachary,adjnoun,football,dolphins,polbooks}.  The radius of the circle is the square root of the largest eigenvalue, which is a heuristic estimate of the bulk of the spectrum. The overlap is computed using the signs of the second eigenvector for the networks with two communities, and using k-means for those with three and more communities. The non-backtracking operator detects communities in all these networks, with an overlap comparable to the performance of other spectral methods.  As in the case of synthetic networks generated by the stochastic block model, the number of real eigenvalues outside the bulk appears to be a good indicator of the number $q$ of communities.  \label{fig:real_spectrum}}
\end{figure}

Finally we turn towards real networks to illustrate the advantages of spectral clustering based on the non-backtracking matrix in practical applications. In Fig.~\ref{fig:real_spectrum} we show $B$'s spectrum for several networks commonly used as benchmarks for community detection. In each case we plot a circle whose radius is the square root of the largest eigenvalue.  Even though these networks were not generated by the stochastic block model, these spectra look qualitatively similar to the picture discussed above (Fig.~\ref{fig:complexplot}).  This leads to several very convenient properties.  For each of these networks we observed that only the eigenvectors with real eigenvalues are correlated to the group assignment given by the ground truth.  Moreover, the real eigenvalues that lie outside of the circle are clearly identifiable. This is very unlike the situation for the operators used in standard spectral clustering algorithms, where one must decide which eigenvalues are in the bulk and which are outside.  

In particular, the number of real eigenvalues outside of circle seems to be a natural indicator for the true number $q$ of clusters present in the network, just as for networks generated by the stochastic block model.  This suggests that in the network of political books there might in fact be 4 groups rather than 3, in the blog network there might be more than two groups, and in the NCAA football network there might be 10 groups rather than 12.  
However, we also note that large real eigenvalues may correspond in some networks to small cliques in the graph; 
it is a philosophical question whether or not to count these as communities.

Note also that clustering based on the non-backtracking
matrix works not only for assortative networks, but also for 
disassortative ones, such as word adjacency networks~\cite{adjnoun}, 
where the important real eigenvalue is negative---without being told which is the case.

A Matlab implementation with demos that can be used to reproduce our
numerical results can be found at \cite{code}.

\section{Conclusion}

While recent advances have made statistical inference of network models for community detection far more scalable than in the past (e.g.~\cite{decelle-etal1,ball-karrer-newman,pseudolikelihood,subsampling}) spectral algorithms are highly competitive because of the computational efficiency of sparse linear algebra.  However, for sparse networks there is a large regime in which statistical inference methods such as belief propagation can detect communities, while standard spectral algorithms cannot.

We closed this gap by using the non-backtracking matrix $B$ as a new starting point for spectral algorithms.  We showed that for sparse
networks generated by the stochastic block model, $B$'s spectral properties are much better than those of the adjacency matrix and its relatives.  In fact, it is asymptotically optimal in the sense that it allows us to detect communities all the way down to the detectability transition.  
We also computed $B$'s spectrum for some common benchmarks for community detection in real-world networks, showing that the real eigenvalues are a good guide to the number of communities and the correct labeling of the vertices. 

Our approach can be straightforwardly generalized to spectral clustering for other types of sparse data, such as real-valued similarities between objects. The definition of $B$ extends to 
\[
B_{(u \to v),(w \to x)} = \begin{cases} 
s(u,v) & \mbox{if $v=w$ and $u \ne x$} \\
0 & \mbox{otherwise} \, , 
\end{cases}
\]
where $s(u,v)$ is the similarity index between $u$ and $v$.  As in the case of graphs, we cluster the vertices by computing the top eigenvectors of $B$, projecting the rows of $B$ to the space spanned by these eigenvectors, and using a low-dimensional clustering algorithm such as $k$-means to cluster the projected rows~\cite{clustering-intro}.  
However, we believe that, as for sparse graphs, there will be important regimes in which using $B$ will succeed where standard clustering algorithms fail.  Given the wide use of spectral clustering throughout the sciences, we expect that the non-backtracking matrix and its generalizations will have a significant impact on data analysis.

\begin{acknowledgments}
  We are grateful to Noga Alon, Brian Karrer, Mark Newman, Nati
  Linial, and Xiaoran Yan for helpful discussions.  C.M. and P.Z. are supported
  by AFOSR and DARPA under grant FA9550-12-1-0432.  F.K. and P.Z. have been
  supported in part by the ERC under the European Union's 7th
  Framework Programme Grant Agreement 307087-SPARCS. E.M and J.N. were supported by 
  NSF DMS grant number1106999  and DOD ONR grant N000141110140. 
\end{acknowledgments}

\end{document}